\documentclass[aps,prl,reprint,floatfix,showpacs,longbibliography]{revtex4-1}
\usepackage{graphicx}
\usepackage{amsmath}
\usepackage{amssymb}
\usepackage{amsfonts}
\usepackage{bm}        % for math
\usepackage[mathscr]{eucal}
\usepackage[colorlinks, linkcolor=red,citecolor=blue,urlcolor=blue]{hyperref}
\usepackage[normalem]{ulem}
\usepackage[all]{hypcap}
\usepackage{multirow}
\usepackage{nicefrac}
\usepackage[dvipsnames,usenames,table]{xcolor}

\begin{document}
\title{Temporal order in periodically driven spins in star-shaped clusters}
\author{Soham Pal, Naveen Nishad, T S Mahesh, G J Sreejith}
\affiliation{Indian Institute of Science Education and Research, Pune 411008, India}
\begin{abstract}
We experimentally study the response of star-shaped clusters of initially unentangled $N=4$, 10 and 37 nuclear spin-$\frac{1}{2}$ moments to an inexact $\pi$-pulse sequence, and show that an Ising coupling between the centre and the satellite spins results in robust period-two magnetization oscillations. The period is stable against bath-effects but the amplitude decays with a time scale that depends on the inexactness of the pulse. Simulations reveal a semiclassical picture where the rigidity of the period is due to a randomizing effect of the Larmor precession under the magnetization of surrounding spins. The time scales with stable periodicity increase with net initial magnetization even in the presence of perturbations, indicating  a robust temporal ordered phase for large systems with finite magnetization per spin.
\end{abstract}
\maketitle
Spontaneous symmetry-breaking is a central notion in many body physics, allowing us to explain  several natural phenomena such as formation of a magnet or ice crystals. While there are many systems in which the underlying spatial symmetries are broken resulting in various crystalline phases, and a few classical systems that exhibit spontaneous temporal oscillations, it was only recently that the possibility of spontaneous breaking of time translation symmetry in quantum systems was considered. The initial proposals \cite{WilczekA2012} for realizing a spontaneous breaking of continuous time translation symmetry were later shown to be forbidden in static equilibrium systems \cite{Bruno2013,Watanabe2015}. However, in the attempt to understand quantum thermodynamics of driven systems, it was realized that an externally driven, disordered, interacting spin system can stabilize a phase which spontaneously break the discrete time translation ($\mathbb{Z}$) symmetry of the system to a subgroup $n\mathbb{Z}$ \cite{Khemani2016,Else2016,Keyserlingk2016,Yao2017}. The phenomenon was soon experimentally realized in trapped cold-atom systems that mimic a long range interacting disordered spin-half chain \cite{Zhang2017}, and in dense collections of randomly interacting nitrogen vacancy spin impurities embedded in diamond \cite{Choi2017,PhysRevLett.119.010602}. While this work was under review, similar observations were also realized in other solid NMR experiments\cite{rovny2018observation}.
\begin{figure}[h!]
\includegraphics[width=\columnwidth]{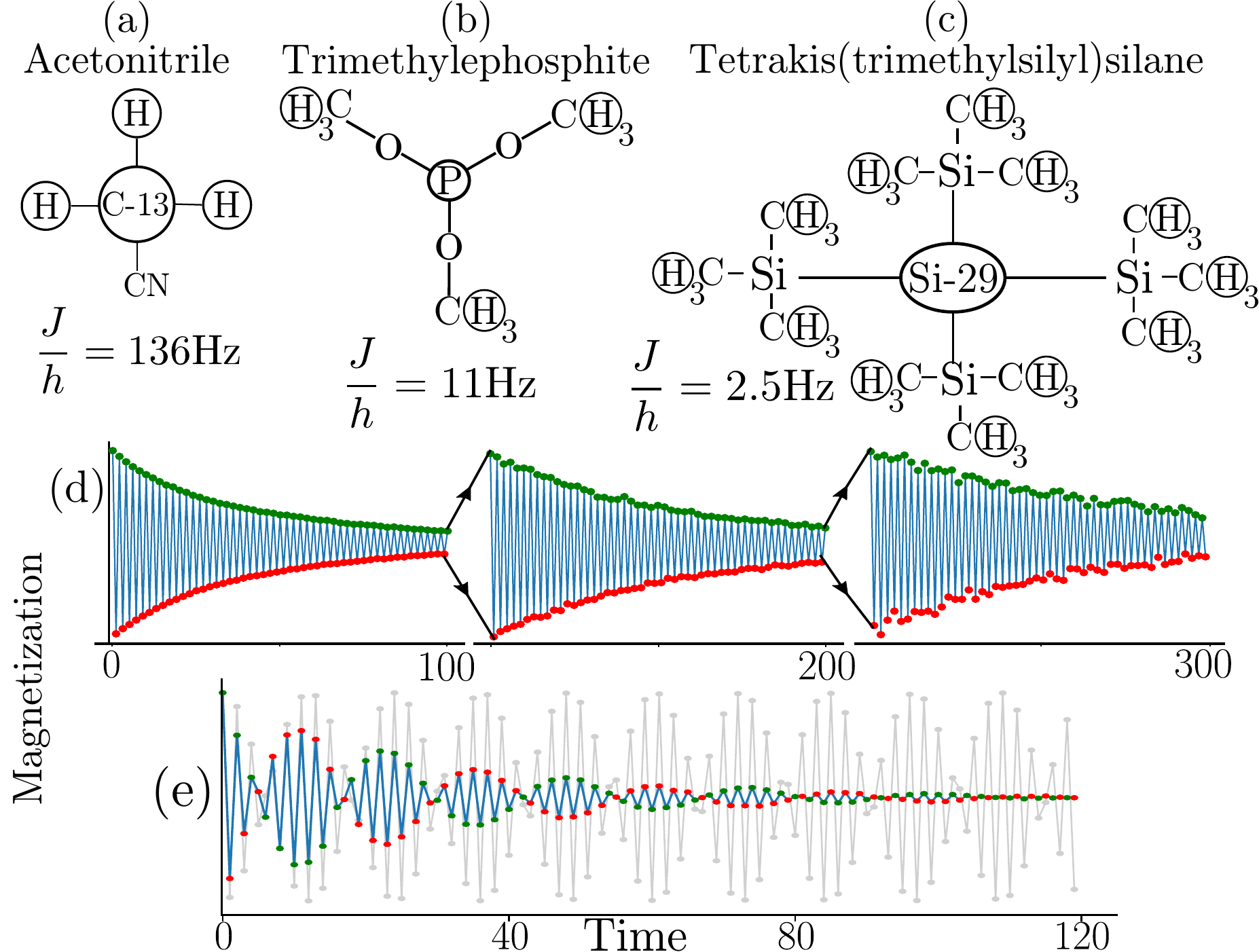}
\caption{Molecules used in the experiments - acetonitrile (a),  trimethyl phosphite (TMP) (b) and tetrakis(trimethylsilyl)silane (c) with the 4, 10 and 37 NMR active nuclei encircled. (d){ Experimentally measured magnetization $\left\langle S_i^z \right \rangle $ of satellite spins of TMP for the pulse sequence in Eq. \ref{Eq-unitary} with $JT/\hbar = 6.5$ and $\theta = \pi-0.1$.} Red/green dots show the magnetization at odd/even time steps. For visibility in the plot, the $y$ axis has been rescaled at every $100^{\rm th}$ time step. (e) Blue line shows experimentally measured magnetization oscillations of free/non-interacting spins of protons in acetonitrile which contain a spinless ${\rm C-12}$ central spin, at a pulse angle $\theta=\pi-0.27$. Gray lines indicate the expected response in the absence of a bath.
\label{molecules}}
\end{figure}

In this work, we report on the observation of robust period two oscillations of magnetization in a cluster of nuclear spins in a simple star-shaped geometry with a central spin interacting with $N$ surrounding satellite spins via Ising interactions mediated by the electron cloud in the molecule. The satellite spins do not interact with each other.
Spins in each molecule show magnetization oscillations of period-two, as expected, when subjected to a sequence of transverse $\pi$-pulses (pulses that rotate every up/down spin by $\pi$ radians). However the Ising interactions within the cluster result in the period rigidly locking on to two, even under a sequence of inexact $\pi$ pulses (pulses that rotate by an amount $\pi-e$). Simulations of an isolated cluster show that the period is robust even in the presence of small perturbations and disorder that break the symmetries of the model. 
For the present work we perform nuclear magnetic resonance (NMR) experiments on acetonitrile, trimethyl phosphite (TMP) and tetrakis(trimethylsilyl) silane (TTSS) containing $4,10$ and $37$ spins [Fig \ref{molecules} (a-c)] \cite{Pande2017}. The experiments are performed on ensembles of $\sim 10^{15}$ molecules with a distribution of initial states, described by a direct product density matrix. High precision ensemble average magnetization measurements of central/satellite spins can be performed using free-induction decay signals. Period-two oscillation of individual spins result in corresponding oscillations of the ensemble average magnetization.
Control experiments performed on molecules that contain a spinless isotope at the center show oscillations with frequencies that linearly vary with the deviation $e$, showing that the robustness of the period originates from interaction with the central spin. In the following, unless units are made explicit, frequencies are in units where the time period $T=1$.

\paragraph{Model and numerical results}:
The unitary operator evolving the state of the cluster between successive steps is given by 
\begin{gather}
U\left(J,\theta;t\right)=\exp\left[-\frac{\imath Jt}{\hbar}S_{0}^{z}{\textstyle\sum_{i=1}^{N-1}}S_{i}^{z}\right]\text{ for }t\in[0,T)\label{Eq-unitary}\\
U\left(J,\theta;T\right)=\exp\left[-\imath\theta{\textstyle \sum_{i=0}^{N-1}}S_{i}^{x}\right]\exp\left[-\frac{\imath JT}{\hbar}S_{0}^{z}{\textstyle \sum_{i=1}^{N-1}}S_{i}^{z}\right]\nonumber
\end{gather}
where $J$, $T$ and $\theta$ are the Ising interaction strength, time period and the rotation angle characterizing the pulse. $S_i^\mu$ are spin operators. Site index $i=0$ labels the central spin (See Ref [\onlinecite{levittbook},\onlinecite{footnote1}] for a description of liquid-state NMR which realizes the unitary)

\begin{figure}
\includegraphics[width=.9\columnwidth]{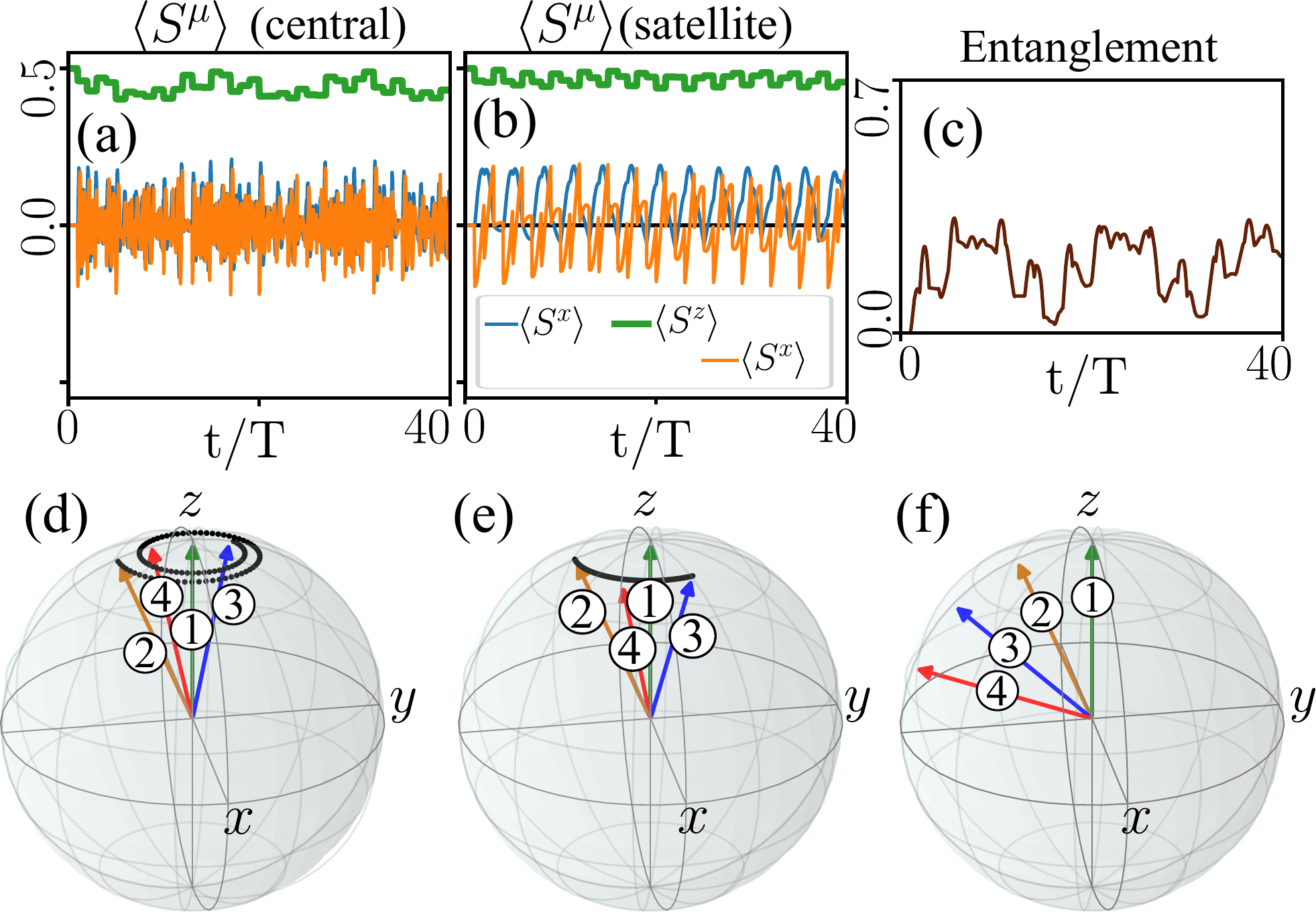} \caption{Numerical simulations of spins in the rotating basis. (a-b) Time dependence of the expectation values of the three spin components of a central (a) and satellite (b)  spin for a system with $N=8$ spins, $\frac{JT}{\hbar}=4$ and pulse angle $\theta=e=0.4$. Initial state is the fully $z$-polarized state. (c) Entanglement entropy of the central spin. (d) Bloch sphere representing the spin components of a central spin (of a $6$ spin cluster) at times $t=0$, $T^+$, $2T^-$ and $2T^+$; $+/-$ labels the time just after/before the pulse. Sequence of intermediate dots track the evolution between time $t=T$ and $2T$. (e) Same as (d) but for a satellite spin. (f) Bloch vectors for a single isolated spin at successive time steps.
\label{Fig-blochcomponents}}
\end{figure}

We will label the deviation from $\pi$ pulse by $e=\pi-\theta$. To simplify the discussion below, it is useful to temporarily switch to a toggling frame of reference in which the basis of every spin rotates by an angle $\pi$ about the $x$-axis after each pulse. On account of the $\mathbb{Z}_2$ symmetry of the model, the unitary operator in the rotating basis retains the same form as in Eq. \ref{Eq-unitary} but with a reduced pulse angle $e=\pi-\theta$, {\it i.e.,} the spins in the rotating basis see a unitary operator $U(J,-e;t)$. A constant $z$-magnetization of all spins in the rotating basis picture corresponds to a period-two oscillation of all physical spins. Numerical simulations indeed show that a finite magnetization is maintained under a sequence of weak pulses (pulse angle $-e$). Presented below is a semiclassical picture inferred from numerical simulations (Fig \ref{Fig-blochcomponents}).

For simplicity, we will consider the time evolution starting from a fully polarized initial state under a sequence of small pulses $\theta=-e$ (corresponding to $\theta=\pi-e$ experienced by the physical spins). During $0<t<T$, the spins do not evolve as the state is an eigenstate of the unitary evolution (Eq. \ref{Eq-unitary}). At time $t=T$, the pulse rotates every spin by an angle $e$ away from $z$-axis as shown on the Bloch sphere (see Fig-\ref{Fig-blochcomponents}). During $T<t<2T$, the central spin which is tilted away from the $z$-axis evolves under the Hamiltonian $H\approx -J\left\langle M_{\rm s} \right \rangle S^z_0$ where $M_{\rm s}$ is the net $z$ magnetization of the satellite spins resulting in a Larmor-like precession as shown in Fig-\ref{Fig-blochcomponents}(d). The orientation of the central spin at $t=2T^-$ depends on the amount of precession $~\frac{JT\left\langle M_{\rm s}\right\rangle}{\hbar}$. The $e$ pulse at $t=2T$ now brings the spin vector to a polar angle $0<\theta<2e$. Owing to the precession, the successive $e$ pulses can now cancel each other. In contrast, in a set of non-interacting spins the angles always add constructively leading to a steady increase in the polar angle ($ne$ after $n$ pulses - Fig-\ref{Fig-blochcomponents}(f) ). Thus the randomizing effect of the interaction induced Larmor precession, causes the polarization of the central spin to survive longer than that of an isolated spin. We expect the same effect to be seen also on the surrounding spins except that they precess under the magnetization of the central spin alone resulting in a slower precession of the satellite spins compared to the central spin (Fig-\ref{Fig-blochcomponents}(f)). 
%Figure \ref{Fig-blochcomponents}(a,b) show the components of the Bloch vector $\vec{\left\langle  S \right \rangle}$ as a function of time. 
Constant sign of the Bloch-vector component $\left\langle S^z\right\rangle$ in the rotating basis implies a period two oscillation of the physical spin orientation (Fig-\ref{Fig-blochcomponents}(a,b)).
\begin{figure}
\includegraphics[width=\columnwidth]{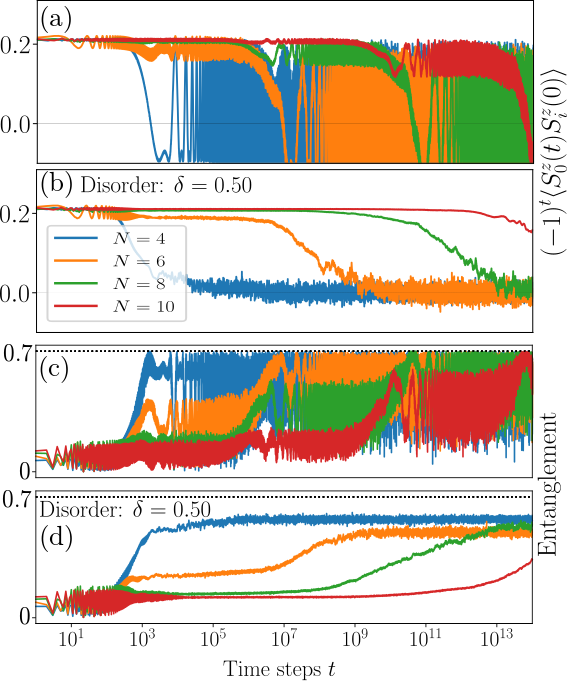}
\caption{(a,b) Time dependence of cross correlation (multiplied by $(-1)^t$) between the central spin $S_0^z$ and a satellite spin $S_i^z$ from simulations of systems of different sizes ($JT/\hbar=4,e=0.05$, $\psi=R_x(\pi/8)\left |\uparrow \uparrow \dots\right \rangle$, $R_x(\pi/8)$ being the rotation of all spins by $\pi/8$ about $x$). Disorder strengths are $0$ (a,c) and $0.5$ (b,d).
(c,d): Entanglement entropy of the central spin. Disorder averaging has been performed in (b,d).
\label{piby8_ent_cc}
}
\end{figure}
Such a semiclassical picture assumes that the central spin is not maximally entangled with the surrounding spins, as otherwise the Bloch vector may vanish in length even when the polar angle is conserved. As shown in Fig-\ref{Fig-blochcomponents}(c), the von Neumann entropy of the central spin stays below maximum ensuring finite Bloch vectors. Simulations of the small systems at much longer time scales using exact diagonalization indicate that entanglement of the system does not rise for time scales that increase exponentially with system size (Fig-\ref{piby8_ent_cc}(c,d)).

In the following, we will use the physical spin basis. To explore the stability of the period to perturbations other than $e$, we numerically simulated a pure spin system with a time independent perturbation to the Hamiltonian of the form $\sum_{i} h^x_iS^x_i + h_i^z S_i^z$. The quenched disorder $h^{x}_i$ and $h^z_i$  were picked uniformly from $[-\delta/2,\delta/2]$ and $[0,\delta]$ (in units where $T/\hbar=1$). To compare the response of different system sizes, we fix the average magnetization per spin. We found that in all cases, the time scale in which there was a dominant period-two oscillation appeared to grow exponentially with system size (Fig-\ref{piby8_ent_cc}(b)). Similar increase in time scales were also observed in simulations with disorder free perturbations of the form $h_z\sum S^z_i$ and $J_x\sum {S^x_iS^x_{i+1}}$\cite{footnote1}. The time scales with stable period are higher when the initial state of the spin cluster had a larger total magnetization.
Slow heating and stability in this disorder free system is likely to be associated with a prethermal regime similar to that in Ref [\onlinecite{ElsePrethermalization},\onlinecite{AbaninSlowHeating}]. However, unlike the high frequency case discussed there, the experiments here are performed at low frequencies ($JT>1$).  Cross correlation between the central and satellite spins (Fig-\ref{piby8_ent_cc}(a,b)) show that different spins oscillate in synchrony suggesting that the robustness of the period is a collective behavior of all spins.

For small $e$ and the $\mathbb{Z}_2$ symmetric unitary (Eq-\ref{Eq-unitary}), origin of the period-two oscillations at finite deviation $e$ can be understood in a manner similar to that described in Ref \cite{Else2016}. The Floquet unitary describing the periodic drive commutes with the parity operator $P=\prod 2S^x_i$ and therefore the quasienergy eigenstates have a parity quantum number $\pm 1$. The quasienergy states of the system at $\theta=0$ occur in degenerate quasienergy pairs of opposite parity $\psi_{\pm} = \left |\sigma_0,m\right \rangle \pm  \left |-\sigma_0,-m\right\rangle$,
where $\left|\sigma_0,m\right\rangle$ is a state with central and satellite spins in an eigenstate of $S_0^z$ and $\sum_{i=1}^{N-1}S^z_i$ with eigenvalues $\sigma_0$  and $m$. At small finite pulse angle $\theta=e$, the quasienergy-degeneracy is broken in a manner that depends on the magnetization $|m|$ as $\sim e^{2|m|+1}$.
In the presence of a sequence of inexact $\pi$ pulses $\theta=\pi-e$, the unitary is $U(J,\pi-e;T)=PU(J,-e;T)$ for which the states $\psi_\pm$ have quasienergies separated by $\pi+\mathcal{O}(e^{2|m|+1})$. A polarized direct product initial state $\left |\sigma_0, m \right\rangle$ is a symmetric or antisymmetric linear combination of the states $\psi_\pm$. As a result, the unitary for inexact $\pi$ pulses acts on such a polarized state to flip the orientation of all the spins at each time step:
\begin{equation}
U\left |\sigma_0,m\right \rangle = U(\psi_+ \pm \psi_-) \sim \psi_+ \pm e^{-\imath \pi}\ \psi_- = \left | -\sigma_0,-m\right \rangle\nonumber
\end{equation}
resulting in a period-two magnetization oscillation. Better degeneracies of the higher magnetization initial states explains why initial states with larger magnetization shows stable periodicity for longer time scales. Subleading ocillations of other frequencies originate from mixing of $\psi_{\pm}$ with states of smaller magnetizations.

\paragraph{NMR setup:} The spin systems used for the experiments - Acetonitrile, TMP and TTSS  are prepared in the solvents dimethyl sulfoxide/deuterated chloroform. The experiments are carried out at $300$ K in a Bruker $400$ MHz NMR-spectrometer equipped with an UltraShield superconducting magnet of strength $9.39$ T. The unitary of Eq-\ref{Eq-unitary} is realized in a doubly rotating frame \cite{levittbook,footnote1}. The $\theta$ pulses are realized by simultaneous resonant, short duration radio-frequency pulses on all spins. The pulse duration can be tuned to control $\theta$. Interaction parameter $JT/\hbar$ can be set by tuning the time period $T$.  After $n$ pulses, any residual transverse magnetization is destroyed using a pulsed-field-gradient (PFG) and the final magnetization $\left\langle S^z \right \rangle$ is rotated into the transverse direction with the help of a $\pi/2$ detection pulse.  The NMR signal is then detected as the oscillatory emf induced in a probe coil due to the precessing transverse magnetization about the Zeeman field \cite{cavanagh,footnote1}. During each period, the measurement was performed immediately after the pulse.

Initial states in the experimental ensemble of $\sim 10^{15}$ molecules can be descibed by mixed state of the form $\rho = \prod_{i=0}^{N-1} \otimes \rho_i$, where $\rho_i=\frac{1}{2}(\mathbb{I}+\epsilon\sigma^z_i)$, and the purity $\epsilon \approx 10^{-5}$, $\sigma^z$ being the Pauli matrix. The purity is inferred from the thermal equilibrium distribution at the magnetic field strength inside the spectrometer. Note that while the ensemble average magnetization is small, the ensemble contains subensembles of all possible initial magnetizations $-N/2\leq M\leq N/2$, with a marginally higher fraction (parameterized by $\epsilon$) with positive sign. Clusters with finite magnetization $|M|$ show stable periodic-two oscillations which collectively reflect in the ensemble average measurements.

\begin{figure}[h!]
\includegraphics[width=\columnwidth]{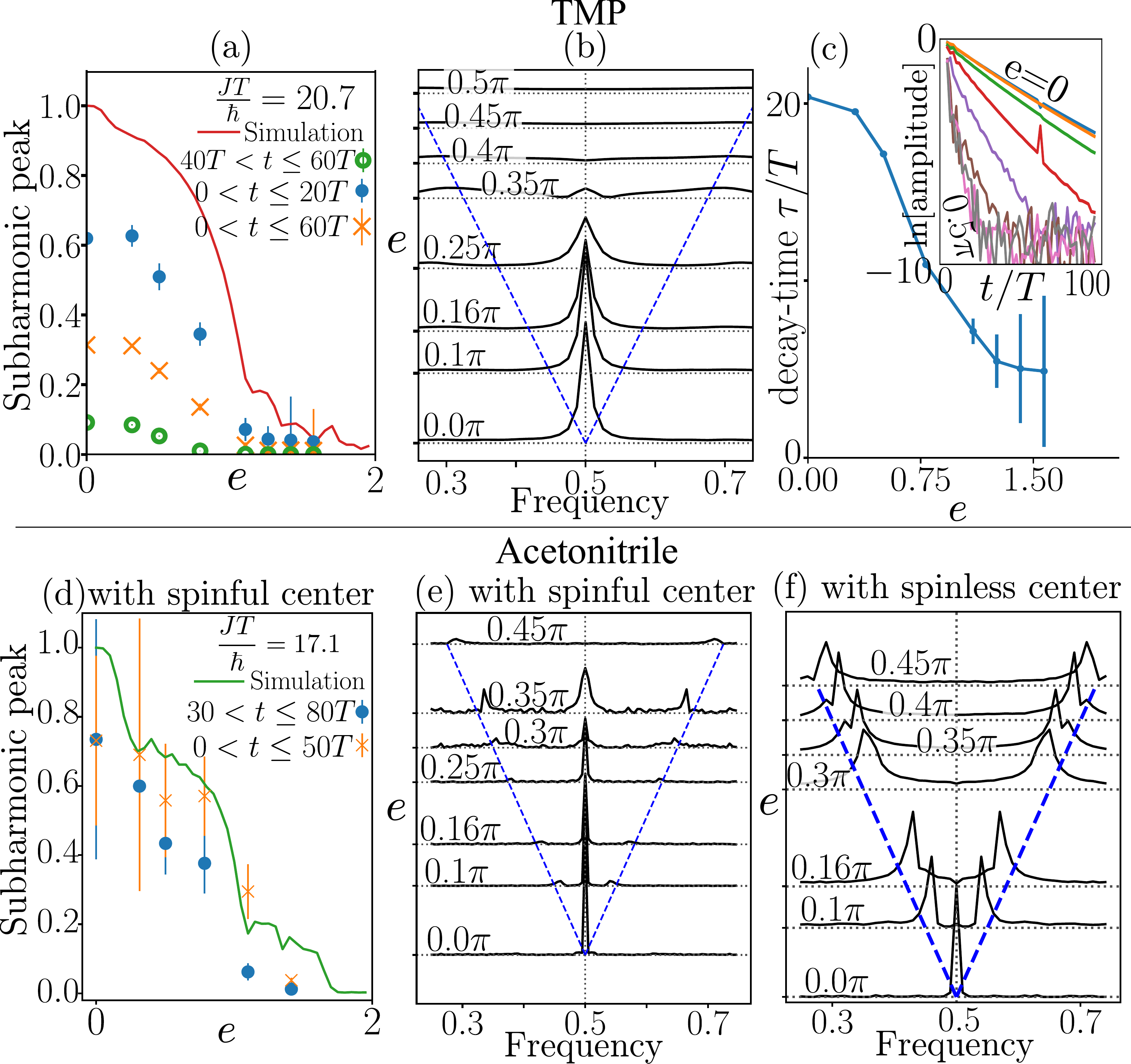}
\caption{ Experimentally measured satellite spin magnetization $ \sum_{i=1}^{N-1} \left\langle S_i^z \right\rangle$. (a,d): Magnitude of the subharmonic peak upon varying $e$ in TMP and Acetonitrile. Solid continuous lines show results from simulations. Different markers indicate Fourier transforms of experimental measurements in different time windows. (b): Waterfall plot of the Fourier spectrum (time-window $0<t<80T$) of the experimentally observerd magnetization of TMP at different deviations $e$. Dashed blue lines indicate the location of peaks expected for a free spin. (c): Variation of the decay time of the experimentally observerd magnetization amplitude with $e$ for TMP. (e,f): Same as (b) but for acetonitrile with a spinfull C-13 (e) and spinless C-12 (f) atom at the center.
\label{Fig:summaryOfSatMeasurements}}
\end{figure}

%%%%%%%%%%%%%%%%%%%%%%%%%%%
\paragraph{Results and discussion}: Fig-\ref{Fig:summaryOfSatMeasurements} shows the measured satellite spin magnetizations in TMP and acetonitrile for an interaction parameter $\frac{JT}{\hbar}=20.7$ ($J/h=11$ Hz, $T=0.3$ s). Magnetization oscillations on TMP (Fig-\ref{Fig:summaryOfSatMeasurements} (a,b,c)) show a clear peak at frequency half (subharmonic peak), whose height decreases with increase in the deviation $e$, vanishing at $e\approx 0.4\pi$ in agreement with the simulations. There are no discernible peaks in the spectrum at frequencies $\frac{\pi\pm e}{2\pi}$ expected from non-interacting spins. Fourier transforms were taken using standard FFT algorithms applied to the data from the chosen time window. For comparability, magnetization data was normalized such that initial magnetization was 1.

\begin{figure}
\includegraphics[width=\columnwidth]{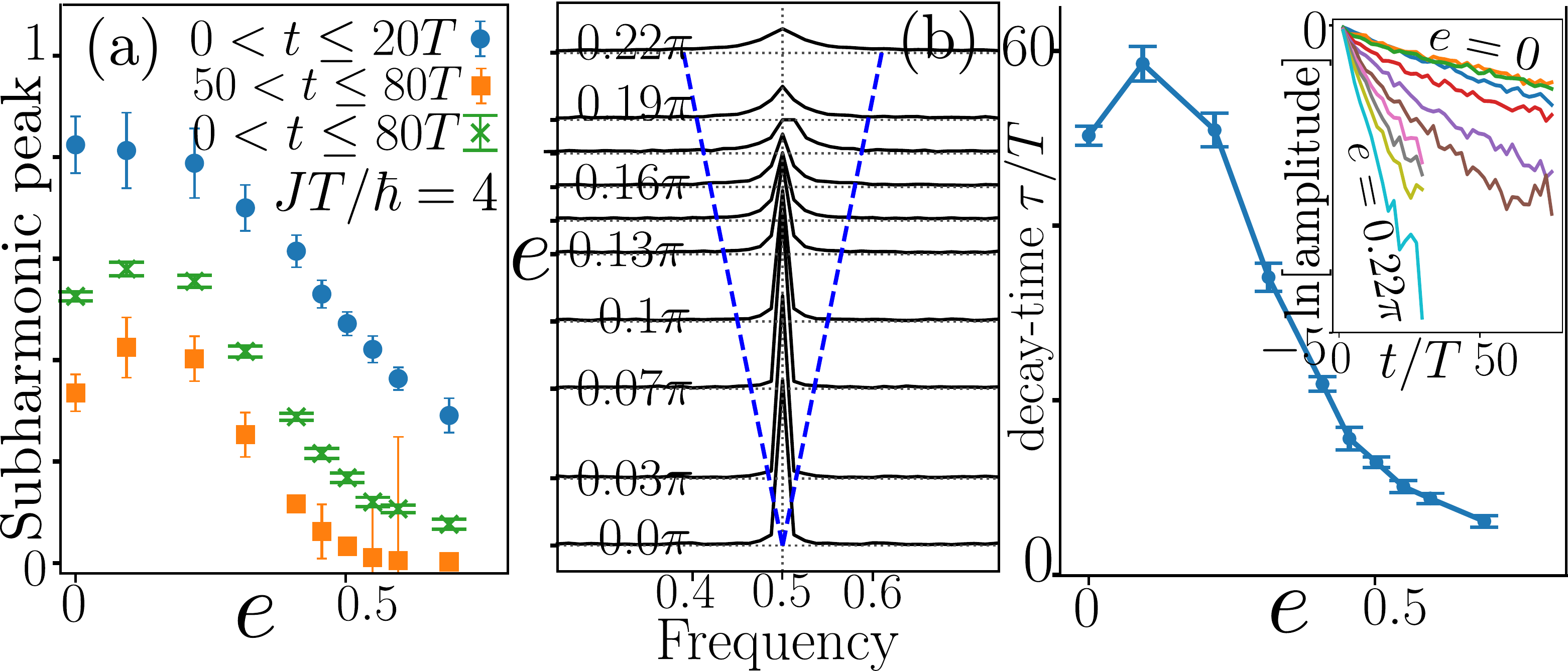}
\caption{
Experimental values of central spin magnetization $ \left\langle S_0^z \right\rangle$ in TTSS. (a): Subharmonic peak strength as a function of the deviation $e$. Different markers indicate Fourier transforms in different time windows. (b): Waterfall plot of the Fourier spectrum of the experimentally observed central spin magnetization at different $e$. Blue dashed line shows the location of the Fourier peaks expected for free spins. (c): Decay time scale as a function of $e$. Inset shows a semi log plot of the amplitude of magnetization as a function of time.
\label{Fig:summaryOfCentralMeasurements}}
\end{figure}
The RF pulses have $\pm 5\%$ distribution of $\mathfrak{\theta}$ values around the nominal value, due to the spatial inhomogeneity of the RF field over the volume of the sample.  The experimental system suffers from decoherence due to coupling to an external thermal bath. This could explain the decay of the oscillation amplitudes with time \cite{PhysRevB.95.195135}. Apart from this decay, the magnitude of the subharmonic peaks in each time window match the simulations. Interestingly the decay time decreases steadily with $e$ (Fig \ref{Fig:summaryOfSatMeasurements} (c)).

%%%%%%%%%%%%%%%%%%

Acetonitrile sample contains a mixture with $99\%$ of the molecules carrying a spinless C-12 and $1\%$ of the molecules containing spinful C-13 atom in the methyl group.  Although NMR signal has contributions from the satellite spins of both isomers, their  contributions can be separated in the frequency domain of the induced emf oscillations during the final measurement process thanks to the presence or absence of interaction with the central spin, and thus they can be analysed separately. Experiments on acetonitrile were performed at the parameter $\frac{JT}{\hbar}\approx 17.1$ ($J/h=136$Hz, $T=.02$s). Figure \ref{Fig:summaryOfSatMeasurements}(e) shows the Fourier transforms of magnetization of the satellite spins in acetonitrile that contain a spinfull C-13 central atom. Figure \ref{Fig:summaryOfSatMeasurements}(f) shows the Fourier transform of the magnetization of the satellite spins in molecules containing a spinless C-12 central atom. In the absence of a central spin with which the satellites can interact, they oscillate like isolated spins with a frequency that varies linearly with $e$. Absence of stable period in this non interacting system clearly shows that the stability of period observed in other clusters arise from interactions. Fig-\ref{Fig:summaryOfCentralMeasurements} shows the results for magnetization measurements of the central Si-29 spin of the TTSS molecule which has $N=36$ satellite spins around the central atom. Experiments were performed at $JT/\hbar\approx 4$ ($J/h=2.5{\rm Hz}$, $T=0.25{\rm s}$). 

We have experimentally demonstrated that stable temporal order can be realized in NMR spin-clusters. Absence of a stable period in the control experiment in C-12 acetonitrile shows that stability of the period requires interactions between the spins (as in C-13 acetonitrile). Though bath effects and other perturbations in the experiment lead to a magnetization decay with time, interestingly the period appears to be unaffected. Stability of the period in the spin cluster improves with increase in total initial magnetization. Therefore large systems with finite initial magnetization per spin, should show a stable temporal ordered phase. The stability of the oscillations in such systems can be interpreted as an error-correction on the pulse sequence and may find potential applications towards robust quantum information processing \cite{choi2017quantum}.

\bibliography{references}

\section*{Supplemental materials}

\subsection*{Effective Hamiltonian of the NMR system}

In this section, we describe origin of the periodic Hamiltonian that governs the dynamics of the spins in each cluster (See Ref-\onlinecite{levittbook} for finer details). 
The spin clusters are contained in each one of about $10^{15}$ molecules dissolved in suitable solvents placed in a high external $z$-directed magnetic field $\mathbf{B}=B_0\hat{z}$. This adds a Zeeman energy term in the Hamiltonian $\gamma _i \mathbf{B}. \mathbf{S}$, where $\gamma _i$ is the gyromagnetic ratio of the $i^{\rm th}$ nuclei. The strong field modifies the electronic environment surrounding the nucleus leading to an additional linear Zeeman-like term $\gamma _i \sum_{\mu,\nu}B_\mu d_{\mu\nu} S^\nu$ where $d_{\mu\nu}$ is called the chemical shift tensor. The shift tensor strongly depends on the type of nucleus and the chemical environment within the molecule. The spins within each molecule interact through pairwise interactions mediated by the electron clouds in the molecule ($J$-coupling), as well as through direct dipole-dipole interactions. Inter molecular interactions are insignificant at the concentrations relevant in this experiment. Thus the Hamitlonian of the system has the form
\begin{multline}
H = \sum_{i} \gamma_i B_0(S_I^z+\sum_\nu d^i_{z\nu}S_{i}^\nu) + \sum_{i<j,\mu,\nu} S_i^\mu J^{\mu\nu}_{ij} S_j^\nu +\\+ H_{\rm dipole} + H_{\rm pulse}
\end{multline}
where the Latin and the Greek indices indicate site and direction indices. $H_{\rm pulse}$ describes the coupling to additional external field that can be used to flip the spins. 

The molecules in the solution have an isotropic environment in which the molecules rapidly rotate. Rotational motion averages the electron mediated spin-spin interactions to its isotropic values whereas the dipole dipole interaction averages to zero (See Chapter-7 of Ref-\onlinecite{levittbook}), reducing the effective interaction to 
\begin{equation}
H = \sum_{\mu,i=0}^{N-1} \gamma_i B_0(\delta_{z\mu}+\bar{d}^i_{z\mu})S_i^\mu + \sum_{i<j=0}^{N-1} J_{ij} {\bf S}_i . {\bf S}_j + H_{\rm pulse}
\end{equation}
where $J$ is the trace of $J_{\mu\nu}$ tensor and $\bar{d}$ is a motionally averaged value of the chemical shift $d$ tensor.

The largest energy scales, Larmor frequency, of the $i^{\rm th}$ nuclei in the system corresponds to the large external magnetic field $B_0$ as, $\omega_i=\gamma_i B_0(1+\bar{d}^i_{zz})$. For proton ($\gamma_i\sim 2.7\times10^8\,{\rm s^{-1}T^{-1}}$) this is of the order of $400 $MHz at $B_0=9.39$T.  In the limit of large spread in $\omega_i$, all those terms that do not commute with the Zeeman term are suppressed. In our case $\vert \omega_0-\omega_i \vert \sim 10^9$ rad/sec, while $J_{ij} \sim 10^2$ rad/sec.  Within this secular approximation, the Hamiltonian simplifies to 
\begin{equation}
H = \sum_{i=0}^{N-1} \omega_i S_i^z + \sum_{i<j=0}^{N-1} J_{ij} S_i^z  S_j^z + H_{\rm pulse}.
\end{equation}
The desired Hamiltonian, which is described in the main text is realized in a rotating frame of reference which rotates at the Larmor precession frequencies of the individual nuclear spins. The time dependent basis transformation that changes the frame of reference $R_i=\exp(-\imath \omega_i S^i_z t)$ commutes with interaction term and therefore in the rotating frame of reference, the Hamiltonian simplifies to
\begin{equation}
H = \sum_{i<j=0}^{N-1} J_{ij} S_i^z  S_j^z + H_{\rm pulse,rot}
\end{equation}
where we have used the time dependent basis transformation for the Hamiltonian $H \to H -\imath \sum_{i}R_i^\dagger\partial_t R_i$. $ H_{\rm pulse,rot}$ is the pulse Hamiltonian expressed in the rotating frame. In order to utilize the external pulse to implement a spin rotation operation, $ H_{\rm pulse,rot}$ should be equal to $\theta_i S_i^x$ (in the rotating frame). This is achieved through a magnetic field pulse that rotates at the Larmour frequency of each nuclear spin and pointed in the $x-y$ plane. With this, the effective Hamiltonian in the rotating frame is 
\begin{equation}
H = \sum_{i<j=1}^{N-1} J_{ij} S_i^z  S_j^z + \sum_{i=0}^{N-1} \theta_i S_i^x
\end{equation}

Note that in the rotating frame of reference, the spins do not see any external magnetic field $B_0$. Since the Hamiltonian is implemented in the rotating frame of reference, the measurement of the $S_z$ expectation values should be performed in the rotating frame. Since the measured quantity $S_z$ commutes with the basis transformation $R$, the actual measurement made in the lab frame is same as the measurement performed in the rotating frame.
 
\subsection*{Measurement and the Pulsed Field Gradient (PFG)}
In order to measure the $S_z$ component of the spins, a detection pulse is applied which rotates all the spins by $\pi/2$ about the $x$-direction, thereby bringing the $z$ component of the spins now to the $x-y$ plane. In the presence of a $z$-directed steady magnetic field, these spin components now precess about the $z$-axis. The collective precession of all the $10^{15}$ spins in the sample can be measured using the EMF induced on a finely calibrated coil. Note that when the $x$-directed detection pulse is applied, any residual component in the $x$ axis remains in the $x-y$ plane, and will then affect the EMF induced in the coil. In order to remove any such $x$ component we apply the pulsed field gradient which randomizes any remnant $x$-components of molecules across the sample.

PFG is a standard NMR technique to create a systematic dephasing of the transverse magnetization along the sample dimension. Typically it involves a linearly varying magnetic field along $\hat{z}$ over a short duration of time $\tau$ (milli-seconds), i.e., $B(z) = B_0 + Gz$. The maximum strength of the gradient $G$ is of the order of 50 G/cm.  The effect of a PFG pulse is to create a phase distribution $\phi(z) = -\gamma B(z) \tau$ in the Larmor precession of the nuclear spins about the $z$-axis such that the average transverse magnetization vanishes. In our experiment, we used a PFG pulse to remove weak residual transverse magnetization, {\emph{if any}}, before tilting the longitudinal magnetization into the transverse plane for detection. The measurement and the PFG system are implemented in Bruker 400 MHz NMR Spectrometer

\begin{figure*}[h!]
\includegraphics[width=\textwidth]{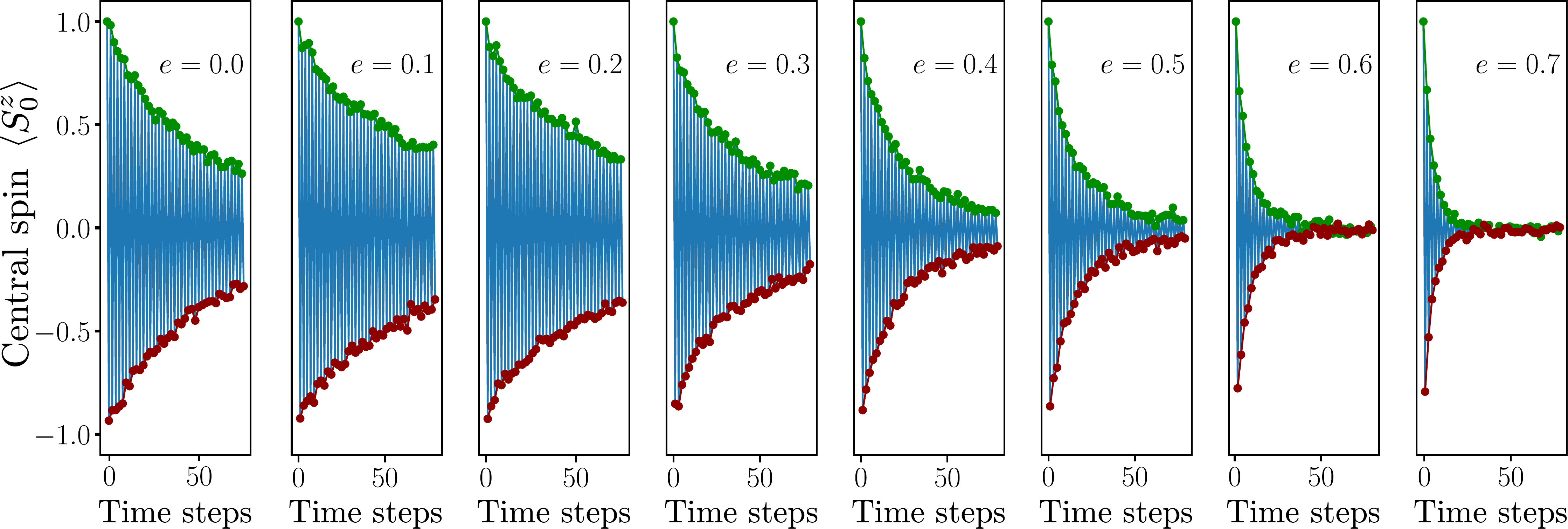}
\caption{Time dependence of the central spin magnetization (obtained from experiments) on the $37$ spin cluster contained in TTSS. Different panels show different values of the deviation $e$ from the exact $\pi$-pulse. The vertical axis has been rescaled such that initial magnetization is normalized to $1$. Green and red dots indicate the magnetizations at odd and even time steps.\label{TTSSdata_time}}
\end{figure*}

\begin{figure*}[h!]
\includegraphics[width=\textwidth]{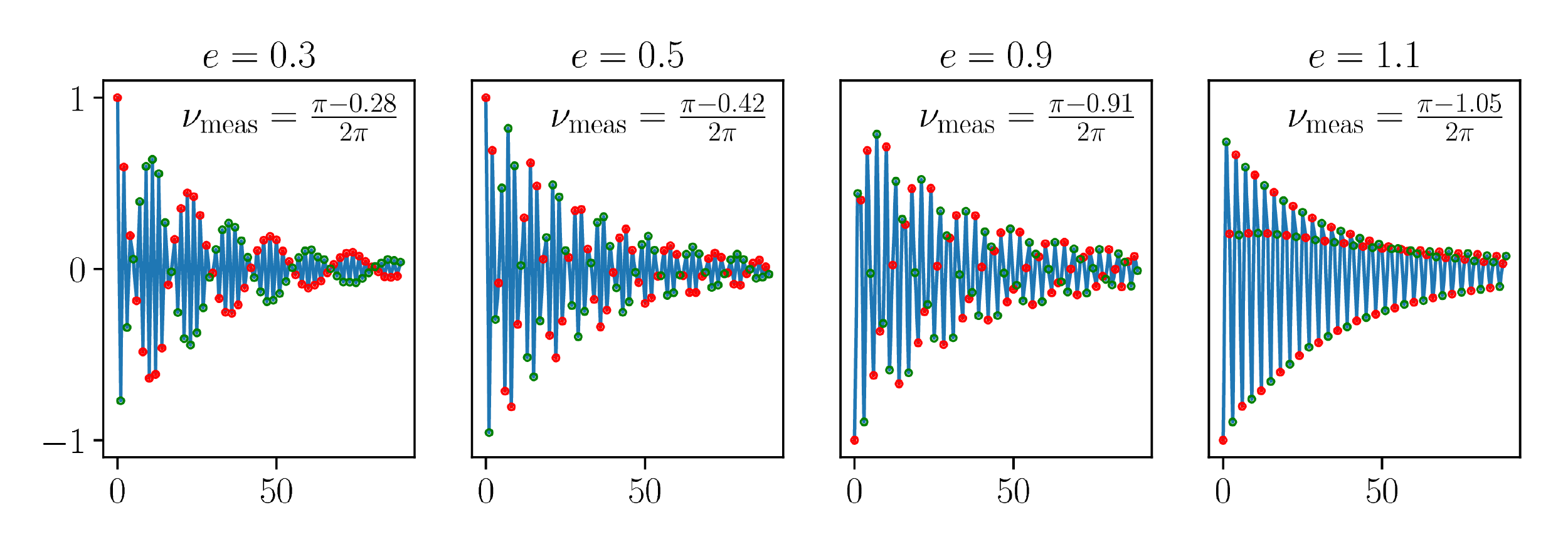}
\caption{Time dependence of the proton spin magnetization (obtained from experiments) in acetonitrile containing a spin-less ${\rm C-12}$ central atom. These spins should behave like free spins which are expected to have oscillations with a frequency of $\frac{\pi-e}{2\pi}$. Different panels show results for different values of the deviation $e$ from the exact $\pi$-pulse. The vertical axis has been rescaled such that initial magnetization is normalized to $1$. Green and red dots indicate the magnetizations at odd and even time steps. Measured oscillation frequencies (peak of the Fourier transform) are indicated in each plot.
\label{C12H3CNdata_time}}
\end{figure*}

\subsection*{Measured magnetization oscillations}

Figure \ref{TTSSdata_time} shows the central spin magnetization of TTSS as a function of timesteps for different deviations from $\pi$ pulse. Figure \ref{C12H3CNdata_time} shows the magnetization of the spins of the protons in acetonitrile molecules which contain a spinless C-12 atom. These spins respond like free spins with magnetization that oscillates with a frequency of  $\frac{\pi-e}{2\pi}$.

\subsection*{Long time simulations of the spins}
\begin{figure*}
\includegraphics[width=\textwidth]{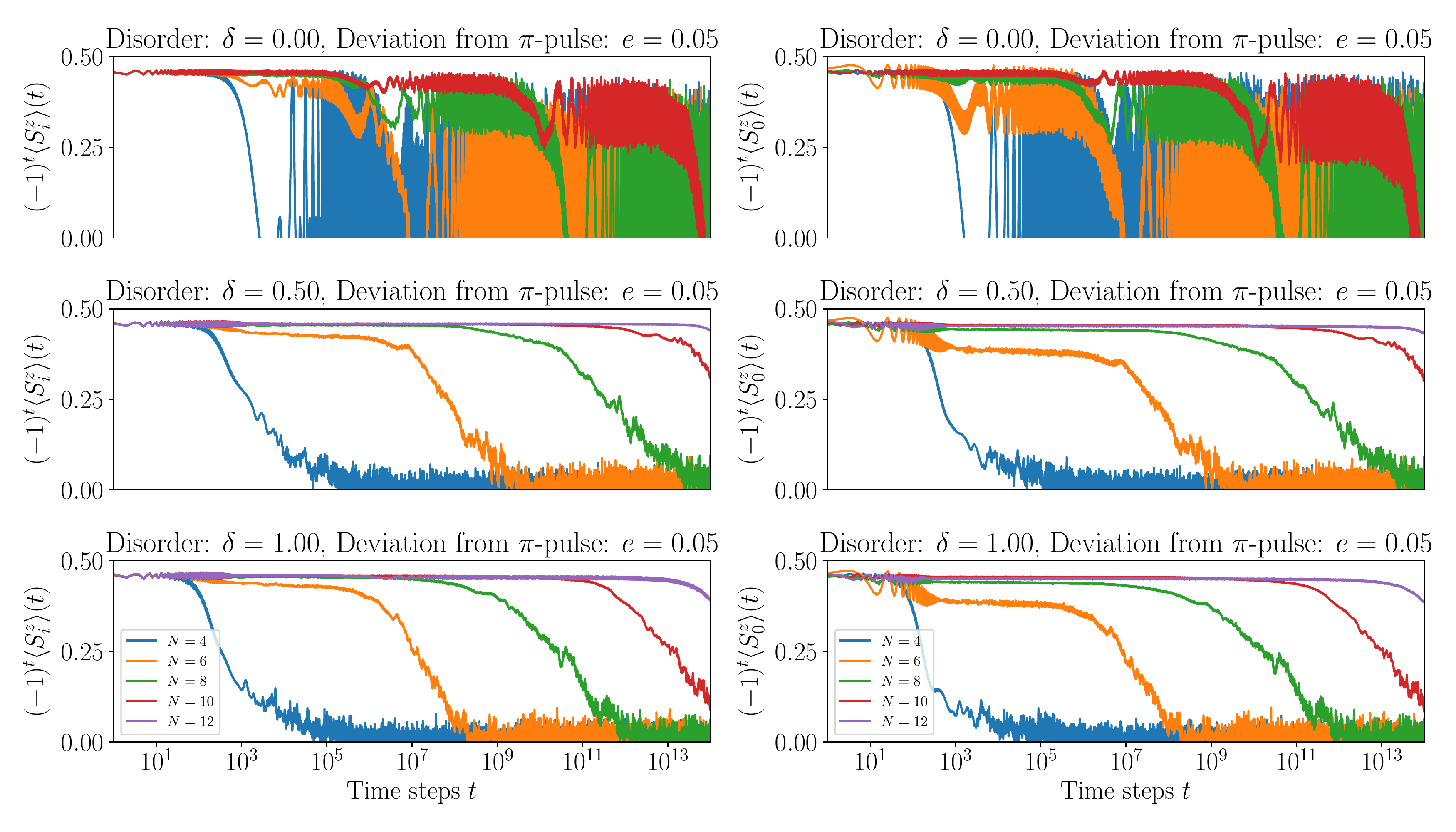}
\caption{Left hand side panels show $\left\langle S_i^z (t)\right\rangle\left(-1\right)^t$ (from simulations) for a satellite spin. Right hand side panels show the same for the central spin. Top, middle and bottom panels show results for different disorder strengths $\delta$ (defined in Eq-\ref{disordermodel}). Deviation from exact $\pi$-pulse is $e=0.05$. Different lines indicate different number of spins $N$. $\frac{JT}{\hbar}=4$ in all the simulations.
\label{piby8_sz}
}
\end{figure*}
Figure \ref{piby8_sz} shows $\left\langle S^z (t)\right\rangle\left(-1\right)^t$ as a function of time for the satellite (left) and central spins (right), starting from an initial state of the form $R_x(\pi/8)\left|\uparrow \uparrow \uparrow ...\right \rangle$, where the rotation operator $R_x(\theta)$ rotates every spin by $\theta$ about the $x$ axis. Top panel shows the time evolution in the absence of any added disorder. Middle and bottom panels show time evolution in the presence of quenched disorder as well as parity symmetry breaking perturbations introduced according to the following model (Similar to the one studied in Ref-\onlinecite{Else2016}):
\begin{align}
&U = U_1 U_2\text{ where }\nonumber\\
&U_2 = 
\exp\left[
\imath \frac{JT}{\hbar} 
S^z_0 \sum_{i=1}^{N-1} S_i^z + \imath \sum_{i=0}^{N-1}  h^z_i S_i^z + \imath \sum_{i=0}^{N-1} h^x_i S_i^x
\right] 
\nonumber\\
&U_1 = \exp\left[-\imath (\pi-e)\sum_{i=0}^{N-1} S_i^x\right]\label{disordermodel}
\end{align}
$h^z_i$ and $h^x_i$ were picked from a uniform distribution in the range $[0,\delta]$ and $[-\delta/2,\delta/2]$ resepectively. For $\delta\neq 0$, the plots show disorder averaged values.

Stable positive value of $\left\langle S^z (t)\right\rangle\left(-1\right)^t$ indicates that there is a dominant period $2$ oscillation. As can be seen in all the cases, when magnetization per spin is held constant, there is a long time scale that appears to grow exponentially with $N$ where this stable period two oscillation is mainitained. The small oscillations in this quantity about the steady value arise from modes at other frequencies. As system size increases, these additional modes weaken. The relatively smooth plots in the case of models with disorder is an artifact of disorder averaging.
\begin{figure*}
\includegraphics[width=\textwidth]{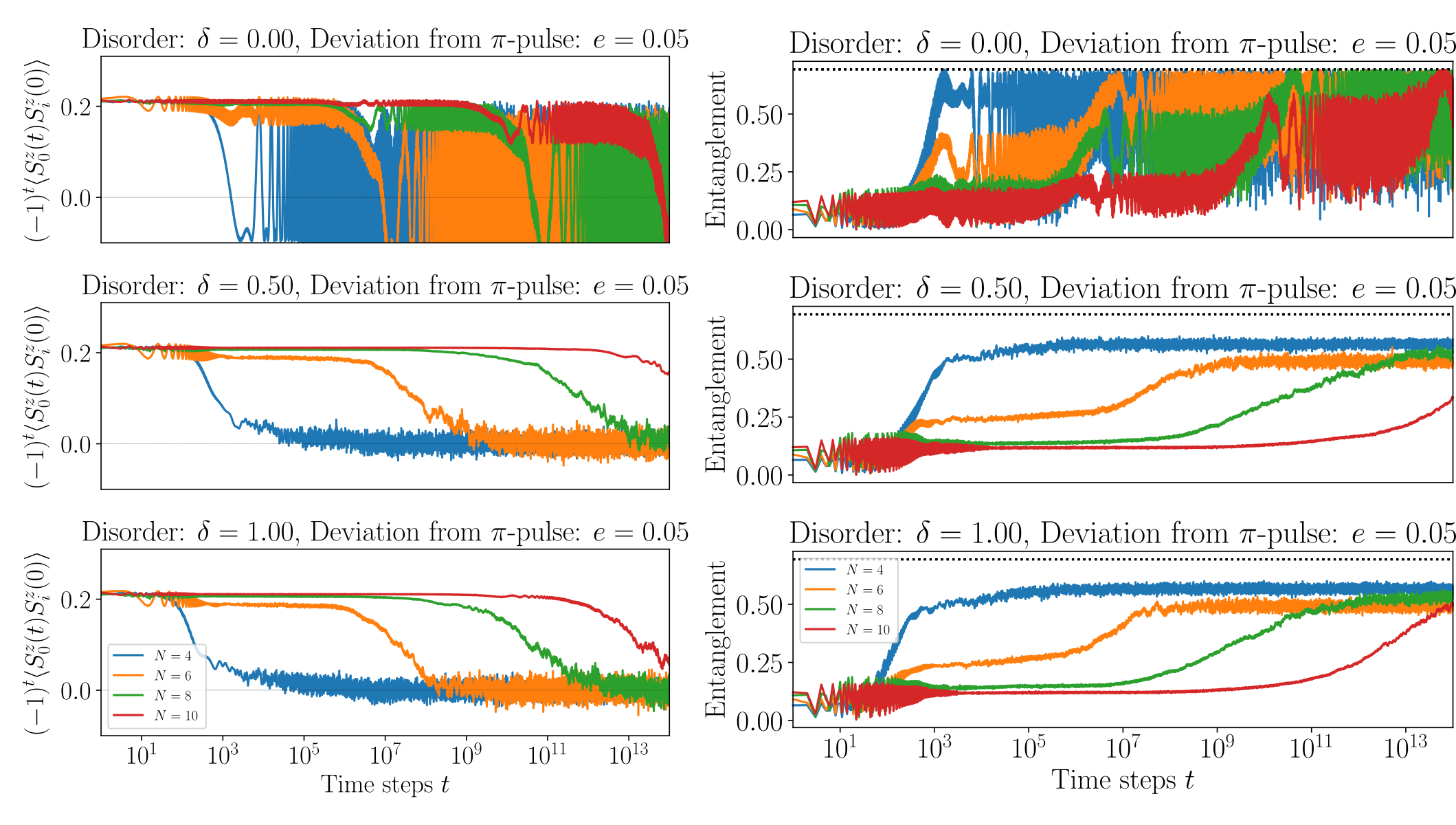}
\caption{(left) Time dependence of cross correlation (multiplied by $(-1)^t$) between the central spin $S_0^z$ and a satellite spin $S_i^z$ at various disorder strengths $\delta=0,0.5$ and $1.0$ and different system sizes at parameter values and initial states same as in Fig-\ref{piby8_sz}. (right)Entanglement entropy of the central spin.
\label{piby8_ent_cc}
}
\end{figure*}

\begin{figure*}
\includegraphics[width=\textwidth]{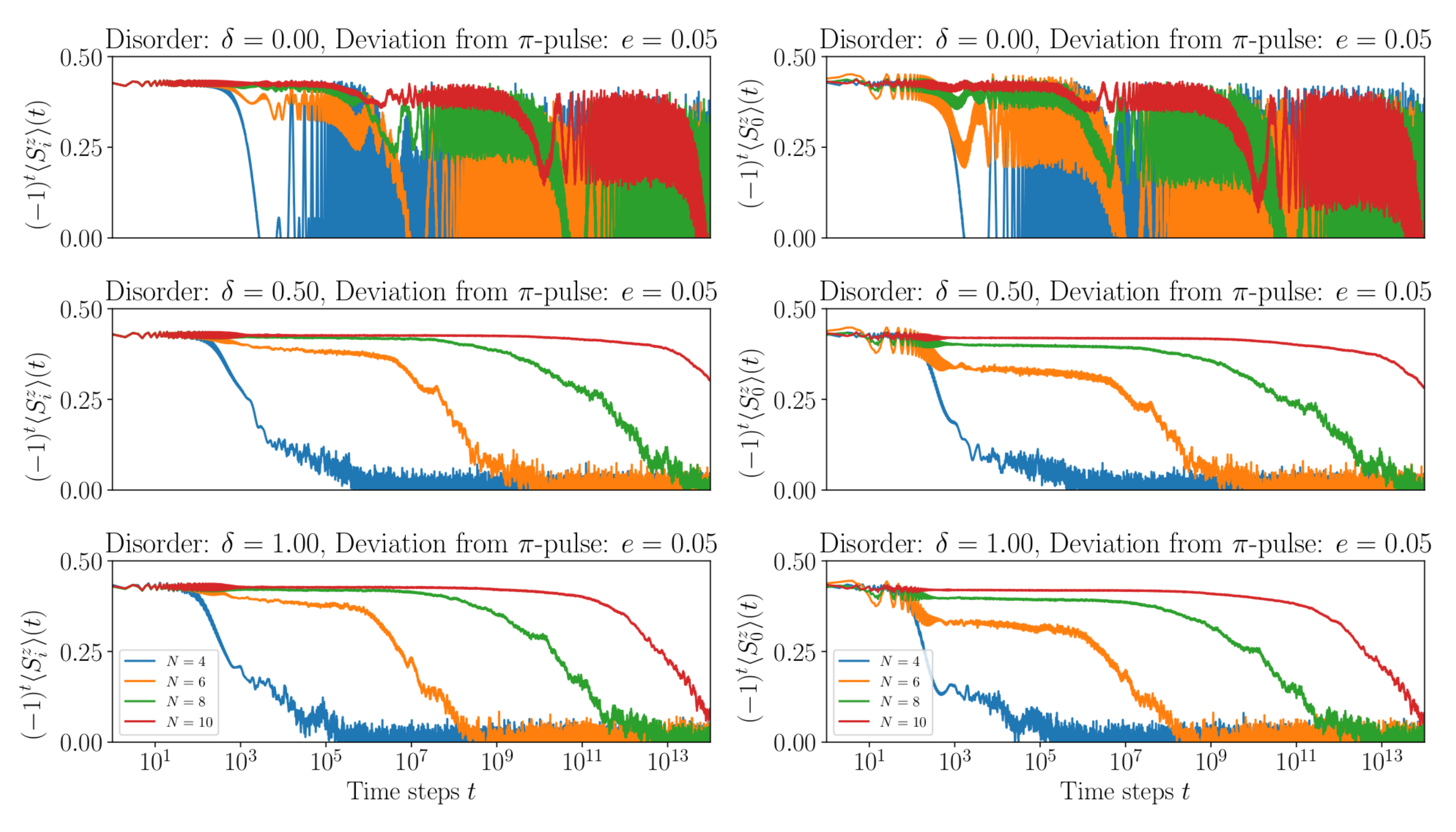}
\caption{Simulations of the spin expectation values at parameters as Fig-\ref{piby8_sz} but with an initiation state with a smaller magnetization $R_x(\pi/6)\left | \uparrow \uparrow \uparrow \dots \right \rangle$
\label{piby6_sz}
}
\end{figure*}

Long-lived finite amplitude period-two oscillations in the spins require that the system do not heat up to an infinite temperature maximally entangled state. We indeed find that the entanglement entropy of the central spin (Fig.\ref{piby8_ent_cc} (right)) do not increase until an large number of time steps, which appears to increase exponentially with system size. Fig.\ref{piby8_ent_cc} (left) panel shows time dependence of the cross correlation between the central and satellite spins at different times showing that the different spins oscillate in a correlated manner. The cross correlation is defined as $\left \langle \psi(0) | U^\dagger(t,0)S_0^z U(t,0)S_i^z|\psi(0)\right \rangle$. In Fig.\ref{piby8_ent_cc} (left) the quantity has been multiplied by the modulating factor $(-1)^t$.

Figure \ref{piby6_sz} shows the results of simulations similar to Fig \ref{piby8_sz} but for an initial state with smaller initial magnetization per spin: $R_x(\pi/6)\left|\uparrow \uparrow \uparrow ....\right \rangle$. While qualitatively the features remain the same, the strength of the subdominant oscillations (of period different from two) increase for a system of same size. However these oscillations vanish as system size increase. Additionally the time scale over which the periodicity is maintained decreases relative to Fig \ref{piby8_ent_cc} upon reducing the initial magnetization per spin.

\begin{figure*}
\includegraphics[width=.95\textwidth]{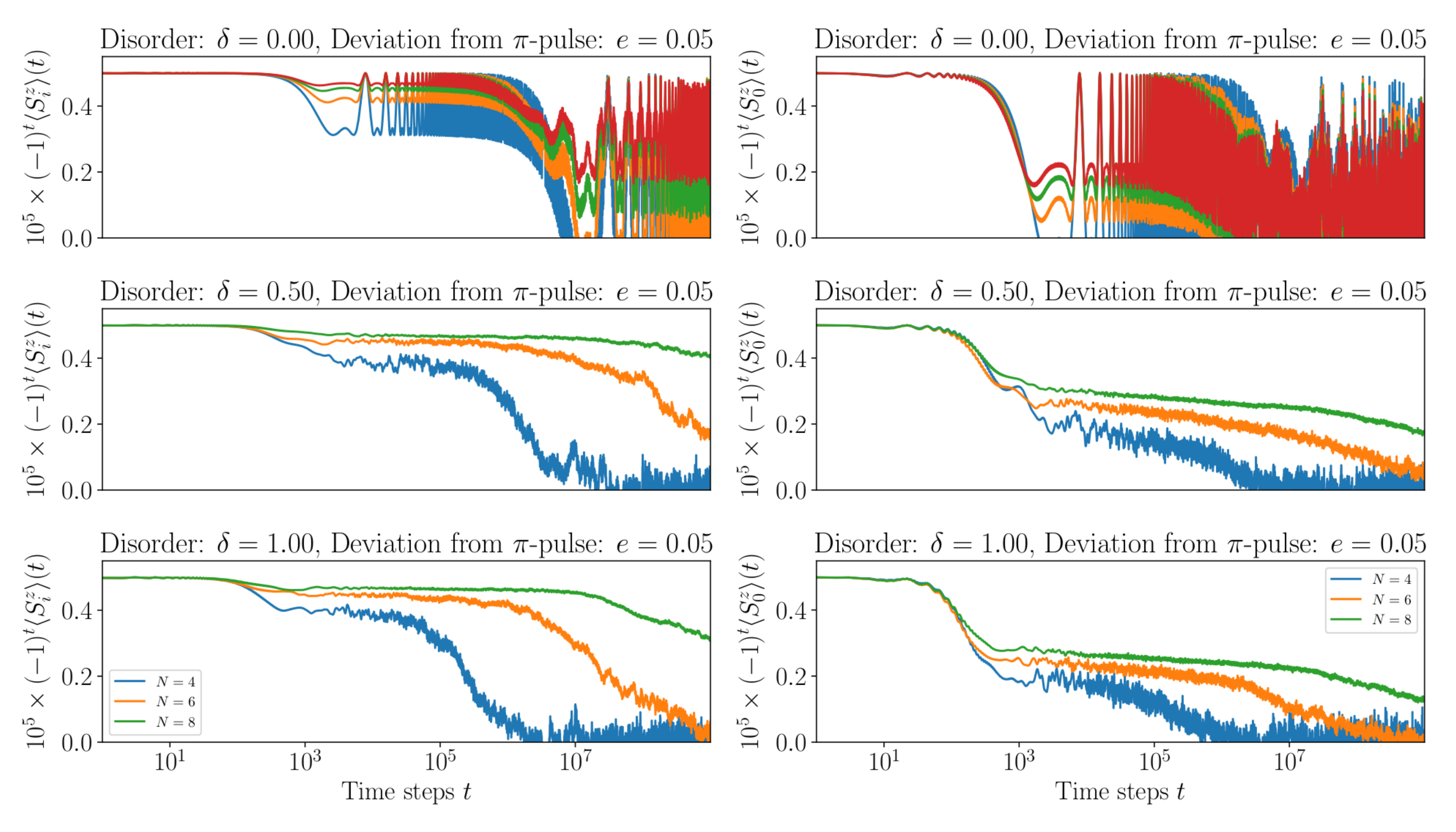}
\caption{Time dependence of ensemble averaged magnetization $\left\langle S^z (t)\right\rangle\left(-1\right)^t$ from simulations of an initial state described by $\rho = \prod_{i=0}^{N-1} \frac{1}{2}[\mathbb{I}_i + \epsilon\sigma^z_i]$ with $\epsilon=10^{-5}$. Interaction strength $JT/\hbar$ and deviation $e$ in the simulation are $4.0$ and $0.05$.
\label{e10_density}
}
\end{figure*}
The experimental setup available realizes a mixed initial state $\rho = \prod_{i=0}^{N-1} \frac{1}{2}[\mathbb{I}_i + \epsilon\sigma^z_i]$ with $\epsilon$ being $10^{-5}$ ($\sigma^z_i$ being the Pauli spin matrix). Even though the ensemble averaged magnetization associated with this is small, the system contains a mixture of cluster states of different initial magnetizations. 
Figure \ref{e10_density} shows the time dependence of $\left\langle S^z (t)\right\rangle\left(-1\right)^t$ for different disorder strengths and deviations from $\pi$ pulse $e=0.05$. We find that the dominant magnetization oscillations survive for time scales far beyond the decay times dictated by the bath. 

\begin{figure*}
\includegraphics[width=\textwidth]{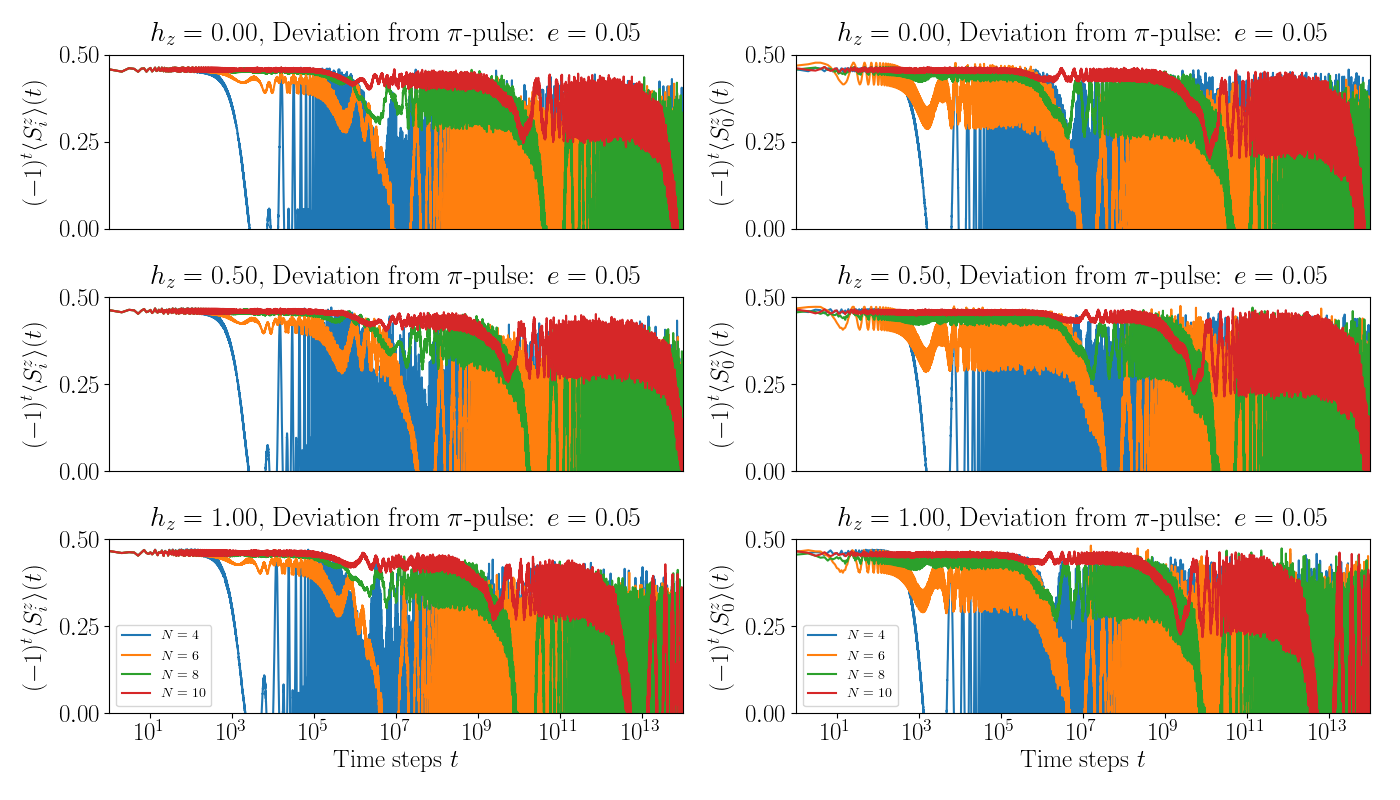}
\caption{Magnetizations estimated from the simulation of a system with a fixed parity breaking perturbation $h_z\sum_{i=0}^{N-1} S^z_i$ and no disorder (Eq \ref{nodisorder_hz}).
\label{piby8_sz_nodisorder}
}
\end{figure*}

Figure \ref{piby8_sz_nodisorder} considers a system without the $Z_2$ symmetry and without disorder described by the unitary
\begin{align}
&U(T,0) = U_1 U_2\text{ where }\nonumber\\
&U_2 = \exp\left[\imath \frac{T}{\hbar} JS^z_0 \sum_{i=1}^{N-1} S_i^z + \imath h^z \sum_{i=0}^{N-1} S_i^z \right] \nonumber\\
&U_1 = \exp\left[-\imath (\pi-e)\sum_{i=0}^{N-1} S_i^x\right]\label{nodisorder_hz}
\end{align}
Initial state in the simulation was same as that in Fig-\ref{piby8_sz} $R_x(\pi/8) \left |\uparrow \uparrow \dots\right \rangle$. We find that the period two oscillations are robust agains $S_z$ perturbations even in the absence of any disorder in the system.

\begin{figure}
\includegraphics[width=\columnwidth]{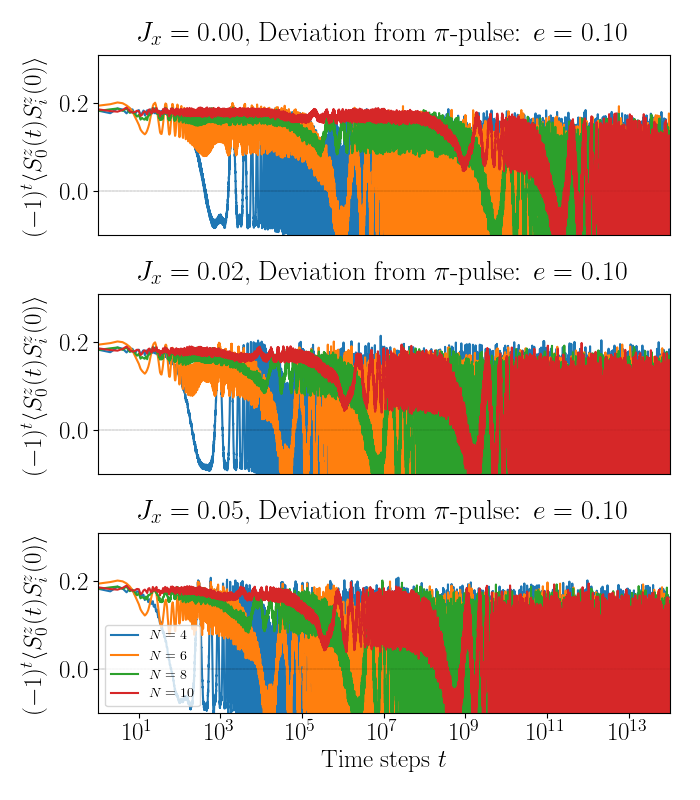}
\caption{Cross correlations from simulations for a with initial state $R_x(\pi/6)\left | \uparrow \uparrow ... \right \rangle$ and pulse deviation $e=.1$ but with an additional perturbation of the form $J_xS_0^x\sum_{i=1}^{N-1}S_i^x$.\label{Jxpert}}.
\end{figure}

Similar studies with an additional interaction term in $U_2$ of the form $U_2 = \exp\left[\imath \frac{T}{\hbar} JS^z_0 \sum_{i=1}^{N-1} S_i^z + \imath J^x S_0^x\sum_{i=1}^{N-1} S_i^x \right]$ shows qualitatively the same behavior (Fig-\ref{Jxpert})

\subsection*{Degeneracy of quasienergies}
In this section we show that for small pulse angle $\theta=e$, the $Z_2$ symmetric unitary
\begin{equation}
U = \exp\left[{-\imath \theta \sum_{i=0}^{N-1}S_i^x}\right]\exp\left[{\imath \frac{JT}{\hbar}S_0^z\sum_{i=1}^{N-1}S_i^z}\right]\label{z2unitary}
\end{equation}
has degenerate pairs of eigenstates, with the degeneracy improving (ie decreasing gaps) with the average magnetization in the eigenstates.

\begin{figure}
\includegraphics[width=\columnwidth]{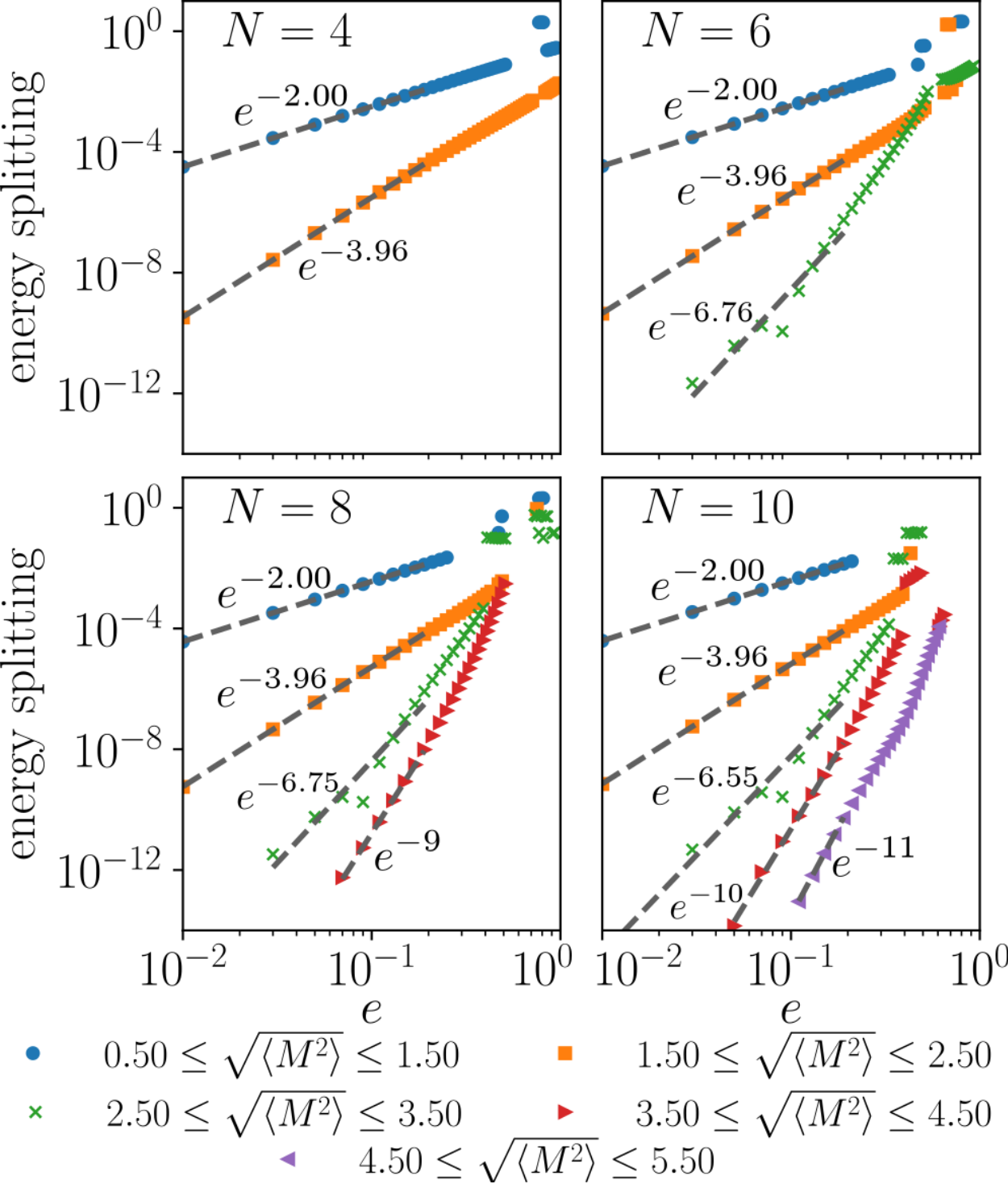}
\caption{Plot showing the average level spacing between quasienergies of opposite parity eigenstates of the Floquet unitary (Eq-\ref{z2unitary}), as a function of the pulse angle.
\label{degeneracy}}
\end{figure}

The eigenstates of the unitary are parity eigenstates. For small $e$, eigenstates are symmetric/antisymmetric linear combinations of localized states opposite magnetization $~\left | +m\right   \rangle \pm \left | -m\right   \rangle$ and therefore eigenstates can be grouped into sets with distinct expectation values $M_{\rm rms}\sim\sqrt{\left \langle M^2 \right \rangle}$. We find that the states with larger magnetizations are more degenerate. To show this we simultaneously diagonalize the parity operator $P$ and the unitary $U$. Eigenstates can be grouped according to $M_{\rm rms}$ of the states (Fig-\ref{degeneracy}). Average spacing between opposite parity quasienergies in a group can be estimated as the average of $|\omega_{i,+}-\omega_{i,-}|$ where $\omega_{i,+/-}$ is the quasienergy of the $i^{\rm th}$ positive/negative parity eigenstate when the states in the group are sorted according to the quasienergy values (taking care of Brillouin zone crossings while evaluating the distance $|\omega_{i,+}-\omega_{i,-}|$).

As shown in Fig \ref{degeneracy}, eigenstates of larger $M_{\rm rms}$ have better degeneracies. The level spacing scales as $e^{2M_{\rm rms}}$. This can be qualitatively understood by treating $U(J,e)$ as a perturbation over $U(J,0)$ of the following form:
\begin{equation}
U(J,e)\approx U(J,0) - \imath e \sum_{i=0}^{N-1}(S_i^x) U(J,0) + \mathcal{O}(e^2)
\end{equation}
For the case of $e=0$, the eigenstates come in degenerate opposite parity pairs $\left|\pm \right  \rangle=\left |+M\right \rangle\pm \left |-M\right \rangle$. At finite $e$, the degeneracy is broken by scattering by the $S_x$ operator. We expect that the lowest order splitting in the degeneracy occurs due to $2M$ actions of the perturbation, leading to a splitting of the unitary eigenvalues of order $e^{2M}$. Quasienergies split similarly as $|\exp[{-\imath \omega}]-\exp[{-\imath (\omega+\delta)}]| \sim |\delta|$ for small $\delta$.

\end{document}